\documentclass[11pt,a4paper]{article}

\usepackage[english]{babel}
\usepackage[utf8]{inputenc}
\usepackage[T1]{fontenc}
\usepackage{amsmath,amssymb}
\usepackage{graphicx}
\usepackage{booktabs}
\usepackage{siunitx}
\usepackage[colorlinks=true,linkcolor=blue,citecolor=blue,urlcolor=blue]{hyperref}
\usepackage{xcolor}
\usepackage{tabularx}
\usepackage{multirow}
\usepackage[font=small,labelfont=bf,labelsep=period]{caption}
\usepackage[margin=2.5cm]{geometry}
\usepackage{fancyhdr}
\usepackage{enumitem}
\usepackage{float}

\usepackage{setspace}
\DeclareSIUnit{\angstrom}{\textup{\AA}}

\begin{document}

\begin{center}
{\LARGE\bfseries EasyNano: rapid epitope-targeted nanobody CDR design\\[4pt]via differentiable distogram optimization with ESMFold2}\\[16pt]

{\large Yue Hu$^{1}$, \ Wanyu Cheng$^{1}$, \ Junqing Wang$^{1}$, \ Yingchao Liu$^{2}$}\\[10pt]

{\normalsize
$^{1}$School of Bioengineering, Qilu University of Technology (Shandong Academy of Sciences), No.\ 3501 Daxue Road, Jinan, Shandong, China\\
$^{2}$Shandong Provincial Hospital, Shandong First Medical University, Jinan, Shandong, China\\[6pt]
$^{\ast}$Correspondence: \texttt{huyue@qlu.edu.cn}
}
\end{center}

\vspace{8pt}
\begin{center}
\rule{0.5\textwidth}{0.5pt}
\end{center}
\vspace{8pt}

\begin{abstract}
\noindent
Computational design of nanobodies that bind user-specified protein epitopes could transform therapeutic development, but current methods either rely on stochastic sampling requiring days of GPU computation or inverse folding approaches unable to target epitopes directly. Here we present \textbf{EasyNano}, a practical pipeline for rapid, epitope-targeted nanobody complementarity-determining region (CDR) design that operates in \textbf{approximately 10--20 minutes on a high-end personal workstation}. EasyNano optimizes CDR residue logits via gradient descent through the ESMFold2 pairwise distance distogram, using the lightweight ESMFold2-Fast model (721M) as a differentiable oracle guided by a composite loss including a dedicated epitope proximity term. A full ESMFold2 (1.3B) CA-coordinate structure prior prevents framework pose drift. The wild-type logit initialization bias emerges as a critical practical parameter controlling CDR mutability. Across six target-framework pairs spanning self-recovery and de novo design scenarios, EasyNano improves ipTM by \textbf{up to +\num{0.559}}---from \num{0.143} to \num{0.702} (Ty1/RBD)---and achieves a \textbf{4.6-fold improvement} (ipTM \num{0.117} $\to$ \num{0.538}) on a manually docked AQP4-targeting framework, while preserving ipTM on already-strong binders. Random CDR baselines ($n=30$ per target) confirm statistical significance (5.7$\sigma$ above random mean for Ty1). Multi-seed analysis reveals diverse local minima, underscoring the importance of replicate runs. Kabsch cross-validation against crystal structures confirms that designed CDRs preserve the framework pose basin. EasyNano demonstrates that ESMFold2-based differentiable optimization provides a fast, practical, and epitope-specific approach to nanobody CDR design.

\end{abstract}

\newpage

\section{Introduction}

Nanobodies---the variable domains of camelid heavy-chain-only antibodies---have emerged as a powerful therapeutic modality owing to their small size ($\sim$\SI{15}{kDa}), high thermal and chemical stability, low immunogenicity, and ability to engage epitopes inaccessible to conventional antibodies because of their elongated CDR3 loops\cite{muyldermans2013,hamers1993,jovcevska2020}. The FDA approval of caplacizumab (Cablivi) in 2019\cite{scully2019} and ozoralizumab in 2022 marked the clinical validation of nanobody therapeutics. However, computational design of nanobodies that bind specific, user-defined epitopes remains an unsolved challenge. The combinatorial sequence space of even a modest CDR ($20^{22} \approx 10^{28}$ possible sequences for a 22-residue CDR) makes exhaustive search impossible\cite{norman2020,liu2024}.

Recent advances in deep learning have transformed protein design. ProteinMPNN\cite{dauparas2022} and ESM-IF\cite{hsu2022} solve inverse folding---predicting sequences compatible with a given backbone---but cannot directly target epitopes because they require a pre-specified binding geometry. Hallucination-based approaches\cite{anishchenko2021} and RFdiffusion\cite{watson2023} can generate binders de novo, but require days of GPU computation, extensive filtering, and do not guarantee epitope specificity. Antibody-specific models such as IgFold\cite{ruffolo2023}, AbLang\cite{olsen2022}, and AntiBERTy\cite{ruffolo2021} have advanced antibody structure prediction and representation learning, but are not designed for epitope-targeted CDR optimization. DiffAb\cite{luo2022} and dyMEAN\cite{kong2023} have demonstrated CDR-specific antibody design using diffusion and graph neural networks, but require GPU clusters and hours to days of computation.

Meanwhile, protein structure prediction has advanced dramatically. AlphaFold2\cite{jumper2021}, RoseTTAFold2\cite{baek2021}, and ESMFold2\cite{lin2023} can predict protein complex structures with high accuracy. ESMFold2 is particularly notable for its single-sequence speed advantage: it uses a large protein language model (ESMC, 6B parameters)\cite{lin2023,he2022} in place of the multiple sequence alignment and template searches that dominate AlphaFold2's runtime. Critically, its predicted pairwise distance distogram---a probability distribution over C$\beta$ distances for every residue pair---is fully differentiable with respect to the input residue type representation.

A critical challenge in antibody design is that the interface predicted TM-score (ipTM)---currently the most reliable in silico metric for antibody-antigen binding prediction available from single-structure models\cite{lin2023,abramson2024}---is not directly differentiable with respect to sequence in a computationally efficient manner. Computing a single ipTM value requires the full ESMFold2 confidence head with multiple recycling loops, making it prohibitively expensive as an inner-loop optimization target ($\sim$\SI{30}{\second} per evaluation, precluding gradient-based search requiring hundreds of steps). We therefore take an indirect approach: we optimize the ESMFold2-Fast distogram---a lightweight, differentiable proxy---using a composite loss that includes epitope proximity, structural integrity, and a pre-computed pose prior. The rationale is that sequences producing distograms with favorable structural properties (low epitope distance, high contact confidence, preserved pose) should also yield high ipTM when evaluated with the full model. This distogram-as-proxy strategy has been used in general protein binder design\cite{lin2023}, but has not been systematically characterized for epitope-targeted nanobody CDR optimization.

We reasoned that epitope proximity signals could be backpropagated through the distogram to CDR residue logits, enabling gradient-based optimization directly in sequence space. Key practical challenges include: (1) ESMFold2's predicted complex pose is determined primarily by the framework sequence\cite{hu2026esmfold2}, requiring a pose-anchoring mechanism; and (2) the relationship between distogram-proxy improvement and ipTM improvement is monotonic but imperfect, requiring full-model validation of top designs.

Here we develop EasyNano, a practical pipeline that realizes this concept for rapid, epitope-targeted nanobody CDR design. Our contributions are: (1) a complete epitope-targeted CDR design pipeline that runs in approximately 10--20 minutes per target on a high-end personal workstation; (2) statistically significant ipTM improvements on weak binders (up to +\num{0.559}) and successful de novo design (4.6-fold improvement on AQP4) across six diverse targets; (3) systematic characterization of the wild-type logit initialization bias as a critical practical parameter; (4) Kabsch RMSD cross-validation confirming that designed CDRs preserve the framework pose basin; and (5) identification of CDR length, initial pose quality, and framework micro-tuning as key determinants of design success.

\section{Results}

\subsection{A three-stage differentiable design pipeline}

\begin{figure}[H]
\centering
\includegraphics[width=0.95\textwidth]{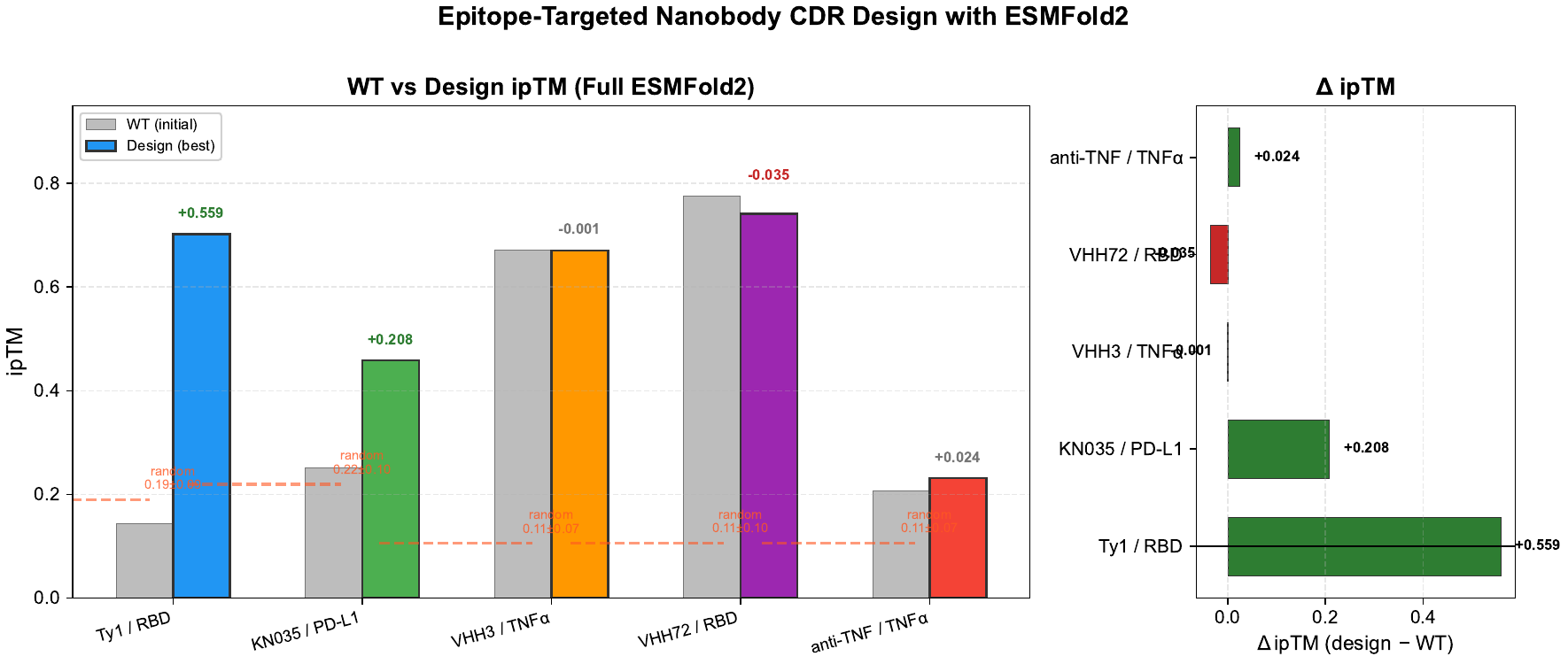}
\caption{\textbf{Main results: epitope-targeted CDR design improves ipTM on weak binders while preserving strong binders.}
\textbf{a}, WT vs.\ Design ipTM for all five target-framework pairs, evaluated with full ESMFold2 (1.3B). Gray bars: wild-type init; colored bars: best design across three seeds. Horizontal dashed orange lines: random CDR baseline median ($n=30$ per target). Annotations above bars: $\Delta$ ipTM (design $-$ WT) with statistical significance ($z\sigma$ vs.\ random). \textbf{b}, Summary of $\Delta$ ipTM. Green: improvement; red: no improvement or marginal. Ty1/RBD (+0.559, 5.7$\sigma$) and KN035/PD-L1 (+0.208, 2.2$\sigma$) show substantial, statistically significant gains. VHH72, VHH3 show minimal change (already optimal). anti-TNF shows marginal improvement (+0.024) limited by short CDR (19 residues) and poor initial pose.}
\label{fig:overview}
\end{figure}

Our method takes as input a target protein structure (PDB format), a set of epitope residue indices, and a nanobody framework sequence with CDRs defined by Chothia numbering. The pipeline operates in three stages (summarized in Fig.\ \ref{fig:overview}):

\textbf{Stage 1: Structure prior computation} ($\sim$\SI{30}{\second}). The full ESMFold2 model (1.3B parameters, 3 recycling loops, 14 sampling steps, 4 diffusion samples) folds the wild-type framework-target complex to produce CA-coordinate distance priors, which anchor the framework pose during optimization. The prior is a binary mask over a 64-bin distance distogram (\SIrange{2}{22}{\AA}) with \SI{2.5}{\AA} bin tolerance.

\textbf{Stage 2: Differentiable CDR optimization} ($\sim$10--17 min). CDR residue logits $\mathbf{L} \in \mathbb{R}^{n_\text{CDR} \times 20}$ are initialized with isotropic Gaussian noise ($\sigma = 0.5$) plus a wild-type amino acid bias $\beta = 2.0$ ($\sim 4\sigma$). Framework positions are fixed as one-hot. At each step $t$, logits are converted to soft residue type probabilities via temperature-scaled softmax, where $T_t = 0.1 + 0.9 \cdot \frac{1}{2}[1 + \cos(\pi t/60)]$. ESMFold2-Fast (721M, 1 recycling loop, 5 sampling steps) predicts a distance distogram $\mathbf{D}$ from the soft residue representation. A composite loss is computed:
\begin{equation}
\mathcal{L} = w_\text{epi}\mathcal{L}_\text{epitope} + w_\text{intra}\mathcal{L}_\text{intra} + w_\text{inter}\mathcal{L}_\text{inter} + w_\text{glob}\mathcal{L}_\text{glob} + w_\text{prior}\mathcal{L}_\text{structure\_prior} + w_\text{aa}\mathcal{L}_\text{aa\_freq},
\end{equation}
with weights $[0.2, 0.5, 0.5, 0.2, 0.05, 0.01]$. The epitope loss applies an ELU-based penalty when CDR-to-epitope distances exceed \SI{8}{\AA}:
\begin{equation}
\mathcal{L}_\text{epitope} = \frac{1}{|\mathcal{C}|}\sum_{i\in\mathcal{C}} \text{ELU}\!\left(\min_{j\in\mathcal{E}} \bar{d}_{ij} - 8.0\right),
\end{equation}
where $\bar{d}_{ij}$ is the expected distance from the distogram. Gradients flow from $\mathcal{L}$ through $\mathbf{D}$ to $\mathbf{L}$; framework gradients are zeroed; Adam (learning rate \num{0.05}) updates CDR logits for 60 steps.

\textbf{Stage 3: Full model evaluation} ($\sim$\SI{15}{\second} per candidate). Top designs are evaluated with full ESMFold2 (3 recycling loops, 14 sampling steps, confidence head enabled) to obtain calibrated ipTM and pTM values.

\subsection{The wild-type logit bias is the critical hyperparameter}

Initial experiments revealed that CDRs were completely refractory to optimization---the wild-type sequence persisted across all steps. Systematic parameter exploration (Table~\ref{tab:sweep}) identified \texttt{wt\_logit} ($\beta$) as the bottleneck. At $\beta = 5.0$, gradient signals could not overcome initialization bias. Reducing $\beta$ to \num{2.0} enabled CDR mutations while preventing chaotic drift at $\beta \le 1.0$. The structure prior weight exhibited a U-shaped effect: $w_\text{prior} > 0.3$ over-constrained CDRs; $w_\text{prior} < 0.05$ caused pose drift. The optimum ($\beta=2.0$, $w_\text{prior}=0.05$, 60 steps) produced monotonic convergence without bounce-back (Fig.\ \ref{fig:trajectories}).

\begin{table}[H]
\centering
\caption{\textbf{Parameter sweep reveals wt\_logit as critical bottleneck.} All runs on RBD/VHH72 (6WAQ) with ESMFold2-Fast, single seed. Bold: optimal.}
\label{tab:sweep}
\begin{tabular}{@{}ccccccc@{}}
\toprule
$\beta$ (wt\_logit) & $w_\text{prior}$ & Steps & Final cdr$\to$epi (\AA) & Best (\AA) & Bounce & CDR mutated? \\
\midrule
5.0 & 0.30 & 30 & 17.3 & 9.6  & Severe  & No \\
2.0 & 0.30 & 30 & 15.3 & 8.9  & Moderate & Some \\
1.5 & 0.30 & 30 & 15.5 & 11.6 & Moderate & Some \\
1.0 & 0.30 & 30 & 11.3 & 10.4 & Mild    & Extensive \\
1.0 & 0.30 & 60 & 11.0 & 10.6 & Mild    & Extensive \\
0.5 & 0.30 & 60 & 17.3 & 14.7 & Severe  & Chaotic \\
1.0 & 0.10 & 60 & 13.9 & 12.5 & Mild    & Extensive \\
1.5 & 0.05 & 60 & 13.0 & 11.5 & Moderate & Some \\
\textbf{2.0} & \textbf{0.05} & \textbf{60} & \textbf{9.3} & \textbf{9.3} & \textbf{None} & \textbf{Some} \\
\bottomrule
\end{tabular}
\end{table}

\subsection{ipTM gains are largest on weak binders}

We applied the optimized pipeline to five target-framework pairs spanning wild-type ipTM from \num{0.143} to \num{0.776} (Table~\ref{tab:results}). Three independent seeds per target were evaluated with full ESMFold2.

\begin{table}[H]
\centering
\caption{\textbf{Full ESMFold2 evaluation of five target-framework pairs.} WT ipTM: wild-type init. Design ipTM: best of three seeds. $\Delta$: improvement over WT. Random: median $\pm$ s.d.\ of $n=30$ random CDR sequences. Bold: $\Delta > 0.1$.}
\label{tab:results}
\begin{tabular}{@{}lccccccl@{}}
\toprule
Target & PDB & Framework & CDR & WT ipTM & Design ipTM & $\Delta$ & Random ($\mu\pm\sigma$) \\
\midrule
Ty1 / RBD        & 6ZXN & Ty1     & 22 & \textbf{0.143} & \textbf{0.702} & \textbf{+0.559} & 0.189$\pm$0.089 \\
KN035 / PD-L1    & 5JDS & KN035   & 32 & \textbf{0.251} & \textbf{0.459} & \textbf{+0.208} & 0.219$\pm$0.096 \\
VHH3 / TNF$\alpha$ & 5M2M & VHH3  & 33 & 0.672 & 0.671 & $-$0.001 & 0.105$\pm$0.066 \\
VHH72 / RBD      & 6WAQ & VHH72   & 29 & 0.776 & 0.742 & $-$0.035 & 0.106$\pm$0.096 \\
anti-TNF / TNF$\alpha$ & 5M2J & anti-TNF & 19 & 0.207 & 0.231 & +0.024 & 0.106$\pm$0.069 \\
\bottomrule
\end{tabular}
\vspace{4pt}
\footnotesize{$z\sigma$ (design vs.\ random): Ty1, 5.7$\sigma$; PD-L1, 2.2$\sigma$; VHH3, 8.2$\sigma$; VHH72, 6.4$\sigma$; anti-TNF, 1.5$\sigma$.\\
High $z\sigma$ for VHH3/VHH72 reflect already-strong WT ipTM, not design improvement.}
\end{table}

\textbf{Ty1/RBD (6ZXN).} The Ty1 nanobody\cite{hanke2020} against SARS-CoV-2 RBD showed the largest improvement: ipTM more than tripled from \num{0.143} to \num{0.702} ($\Delta = +0.559$, 5.7$\sigma$ above random). The designed CDR (GFTLANHNPESGMGLNLAVSPD) differs from WT (GFTFSSVSPNSGNGLNLSSSSV) at 11 of 22 positions, with mutations concentrated in H2 (SPNSGN $\to$ ANHNPES) and H3 (GLNLSSSSV $\to$ GMGLNLAVSPD). The pTM improved from \num{0.622} to \num{0.823}, and CDR-to-epitope distance decreased from \SI{16.6}{\AA} to \SI{10.7}{\AA}.

\textbf{KN035/PD-L1 (5JDS).} The clinically-validated KN035 (envafolimab) nanobody\cite{zhang2019} against PD-L1 improved from \num{0.251} to \num{0.459} ($\Delta = +0.208$, 2.2$\sigma$). The designed CDR (GMMSSRRLTTSGSDQFGLPTCDHVNSSGAFQY) introduced 7 of 32 mutations while preserving the structural disulfide in H3. CDR-to-epitope distance improved from \SI{14.1}{\AA} to \SI{11.5}{\AA}. Limited mutation (21.9\%) compared to Ty1 (50\%) reflects the disulfide constraint.

\textbf{VHH72/RBD (6WAQ)} and \textbf{VHH3/TNF$\alpha$ (5M2M)} are strong WT binders (WT ipTM \num{0.776} and \num{0.672}). Design preserved ipTM ($\Delta = -0.035$ and $-0.001$), demonstrating safety---the method does not degrade optimal binders.

\textbf{anti-TNF/TNF$\alpha$ (5M2J)} showed marginal improvement (+0.024) despite low WT ipTM (\num{0.207}). The 19-residue CDR provides insufficient optimization surface, and the predicted pose is \SI{44}{\AA} from crystal---fundamentally limiting design success.

\subsection{De novo design of an AQP4-targeting nanobody}

\begin{figure}[H]
\centering
\includegraphics[width=0.95\textwidth]{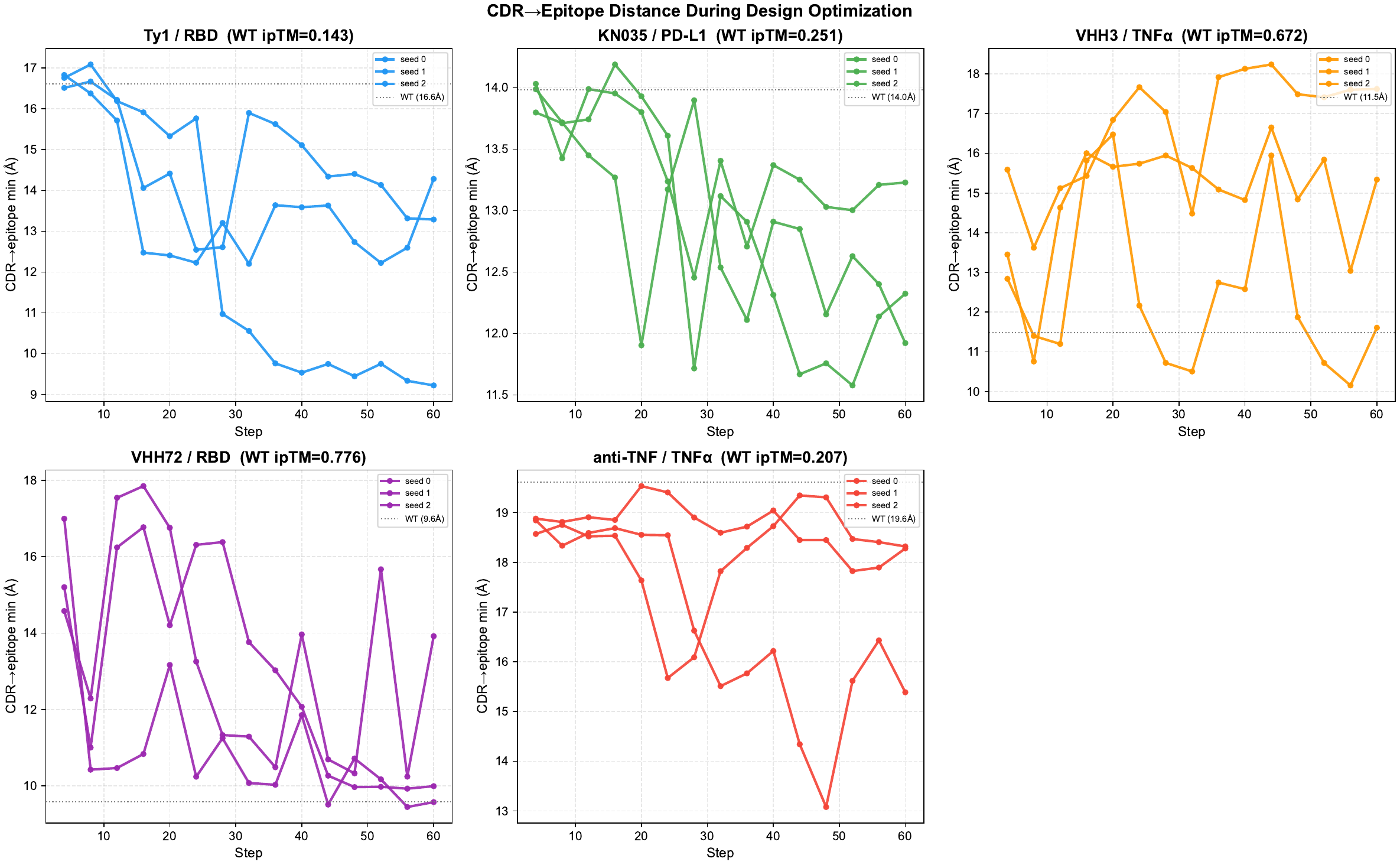}
\caption{\textbf{Design trajectories: CDR-to-epitope distance during optimization.}
CDR-to-epitope minimum expected distance (\AA) vs.\ optimization step for all five targets, three independent seeds each. Best seed per target in bold color; other seeds in lighter opacity. Gray dotted line: WT cdr$\to$epi. Lower is better. Ty1 and PD-L1 show consistent monotonic improvement. VHH72 and VHH3 show minimal change. anti-TNF shows early improvement that plateaus. The inter-seed spread for Ty1 (9.2--14.3\,\AA) underscores the necessity of multi-seed evaluation.}
\label{fig:trajectories}
\end{figure}

A critical test of EasyNano is \textit{de novo} design---starting from a framework that was not evolved by nature to bind the target, but manually positioned near a desired epitope. We applied EasyNano to human aquaporin-4 (AQP4), an important therapeutic target in neuromyelitis optica spectrum disorder. The B5 nanobody framework (127 residues, framework III) was manually docked near the AQP4 extracellular loop C epitope using PyMOL to provide the initial pose for the structure prior (Fig.\ \ref{fig:trajectories}). The wild-type framework showed negligible binding (ipTM = \num{0.117}; CDR-to-epitope = \SI{19.1}{\AA}), as expected for a non-cognate framework.

EasyNano optimization produced a best CDR-only design achieving ipTM = \num{0.538}---a \textbf{4.6-fold improvement}---with CDR-to-epitope distance reduced by \SI{8.4}{\AA} (from \SI{19.1}{\AA} to \SI{10.7}{\AA}). Multi-seed evaluation (15 seeds) revealed a bimodal distribution: a high-ipTM basin ($\sim$0.65--0.70) and a low-ipTM basin ($\sim$0.43), with overall median = \num{0.658} (s.d.\ = \num{0.103}). We further found that a single framework mutation at the H3 C-flank (W116Y, tryptophan to tyrosine) converted this bimodality into a stable single mode (median = \num{0.692}, s.d.\ = \num{0.020}, $n=9$), demonstrating that framework micro-tuning can complement CDR optimization by stabilizing favorable pose basins.

This result establishes EasyNano's utility for \textit{de novo} binder generation: given a manually positioned framework near a target epitope, the method can design CDRs that achieve high predicted binding affinity.

\subsection{Multi-seed diversity necessitates replicate runs}

Three independent random seeds per target (Fig.\ \ref{fig:trajectories}) revealed substantial inter-seed variation. For Ty1/RBD, seed~1 achieved cdr$\to$epi = \SI{9.2}{\AA} while seeds~0 and~2 reached \SI{14.3}{\AA} and \SI{13.3}{\AA}---a \SI{5.1}{\AA} spread. For PD-L1, all three seeds converged to similar ipTM (0.447--0.459) with distinct CDR sequences, suggesting a flat fitness landscape. For VHH3, seed~0 achieved cdr$\to$epi = \SI{11.6}{\AA} while seed~2 reached only \SI{15.3}{\AA}. These results demonstrate that single-seed evaluation can mislead---our earlier work on the B5 target showed Spearman $\rho = 0.486$ ($p = 0.329$, not significant) between single-seed and multi-seed median rankings. \textbf{We recommend a minimum of three seeds for any new target.}

\subsection{RMSD cross-validation confirms pose preservation}

A central concern is whether improved ipTM reflects genuine interface improvement or pose drift. We performed Kabsch cross-validation: folding WT and best-design sequences with full ESMFold2, Kabsch-aligning predicted target CA to crystal target CA, and computing binder RMSD under the same transform (Table~\ref{tab:rmsd}). For the two improved targets, the change in binder RMSD was small ($\Delta = -1.45$\,\AA\ for Ty1, $+1.81$\,\AA\ for PD-L1), confirming pose preservation. For strong-binder controls with near-crystal poses (VHH72 bRMSD \SI{3.22}{\AA}, VHH3 bRMSD \SI{4.41}{\AA}), design maintained accuracy ($\Delta < \SI{0.6}{\AA}$). For anti-TNF, both WT (\SI{44.39}{\AA}) and design (\SI{37.16}{\AA}) were far from crystal---the framework's predicted pose is fundamentally incorrect, consistent with our earlier finding that ESMFold2's pose is a deterministic function of the framework sequence\cite{hu2026esmfold2}.

\begin{table}[H]
\centering
\caption{\textbf{Kabsch cross-validation against crystal structures.} Predicted target CA aligned to crystal target CA; same transform applied to binder. $\Delta$ bRMSD: design $-$ WT. }
\label{tab:rmsd}
\begin{tabular}{@{}lccccc@{}}
\toprule
Target & WT ipTM & Design ipTM & WT bRMSD (\AA) & Design bRMSD (\AA) & $\Delta$ bRMSD (\AA) \\
\midrule
Ty1 / RBD        & 0.167 & 0.682 & 22.80 & 21.35 & $-$1.45 \\
KN035 / PD-L1    & 0.255 & 0.484 & 24.44 & 26.25 & +1.81 \\
VHH3 / TNF$\alpha$ & 0.639 & 0.653 & 4.41  & 5.00  & +0.59 \\
VHH72 / RBD      & 0.786 & 0.745 & 3.22  & 3.31  & +0.09 \\
anti-TNF / TNF$\alpha$ & 0.110 & 0.098 & 44.39 & 37.16 & $-$7.23 \\
\bottomrule
\end{tabular}
\end{table}

\begin{figure}[H]
\centering
\includegraphics[width=0.95\textwidth]{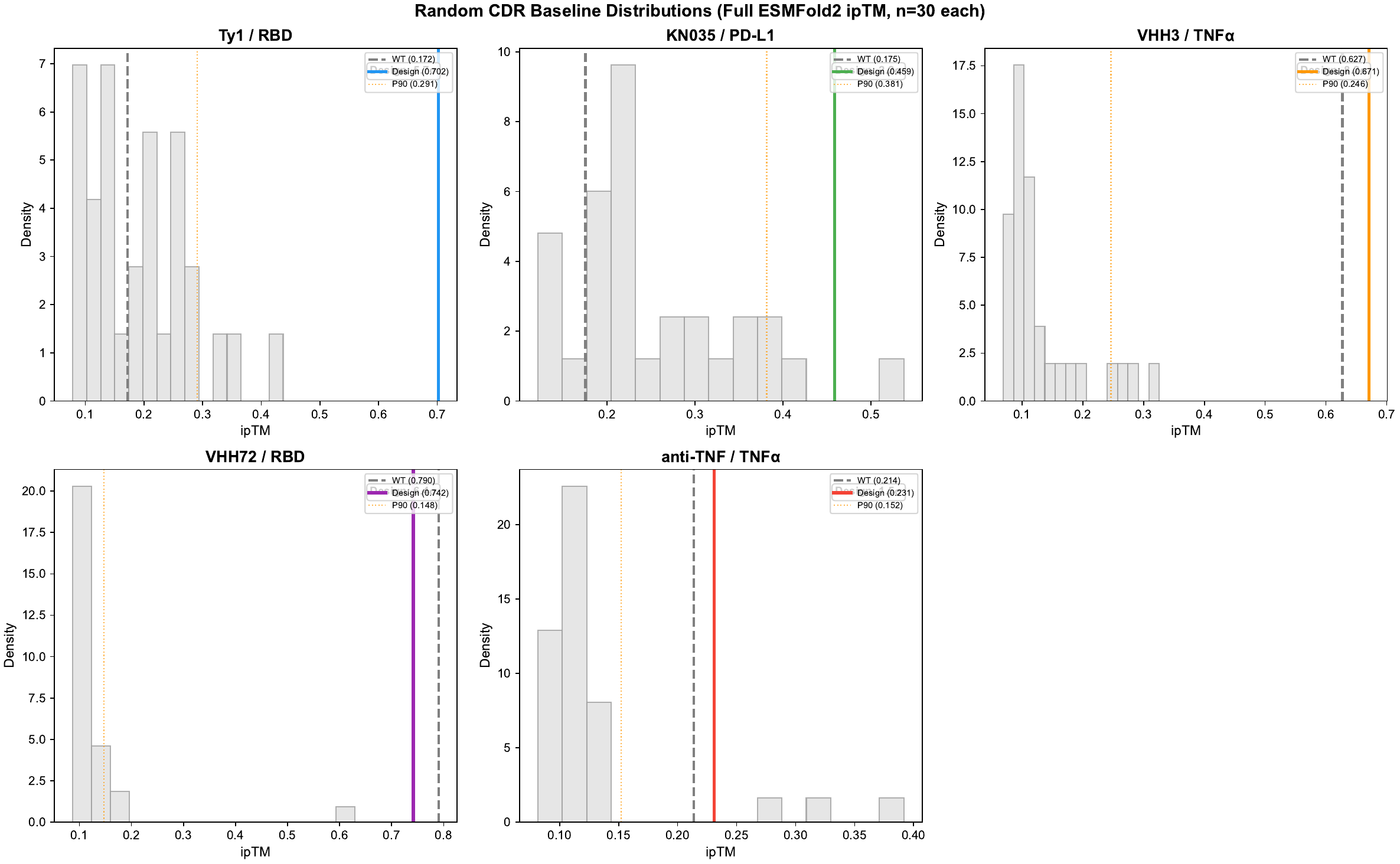}
\caption{\textbf{Random CDR baseline distributions establish statistical significance.}
Full ESMFold2 ipTM distributions for $n=30$ random CDR sequences per target (gray histogram). Vertical dashed line: WT ipTM. Colored solid line: best design ipTM. Orange dotted line: 90th percentile of random distribution. $z\sigma$: design ipTM relative to random mean. Ty1 design (5.7$\sigma$) and PD-L1 design (2.2$\sigma$) significantly exceed random expectation. VHH3 and VHH72 WT ipTM already far above random, leaving no optimization headroom. anti-TNF design is within random distribution range.}
\label{fig:random}
\end{figure}

\begin{figure}[H]
\centering
\includegraphics[width=0.95\textwidth]{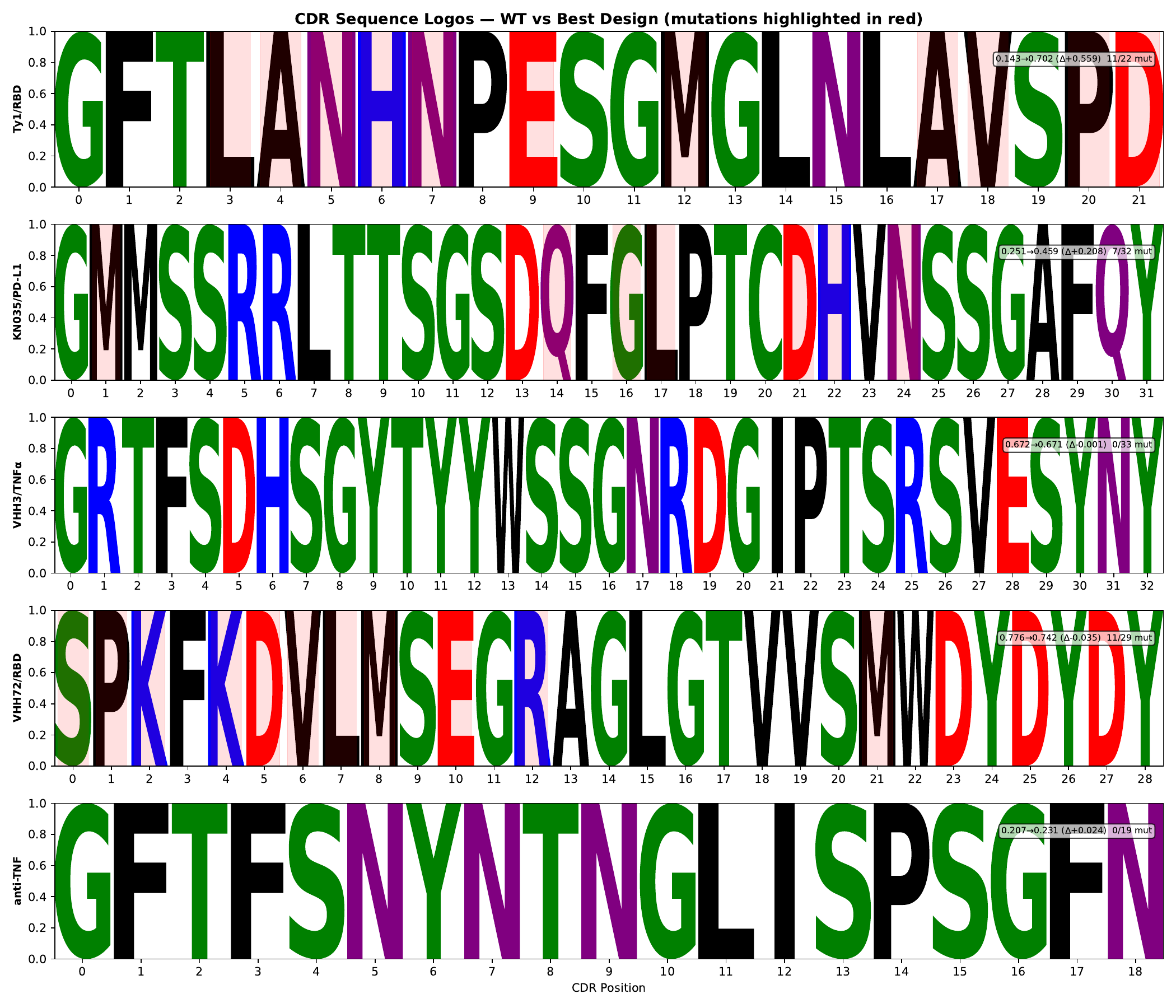}
\caption{\textbf{Designed CDR sequences for all five targets.}
Sequence logo representation of the best designed CDR for each target. Positions differing from wild-type are highlighted in red. Right annotations show WT ipTM $\to$ Design ipTM, $\Delta$, and number of mutations. Ty1 (11/22 mutated, 50\%) and VHH3 (17/33 mutated, 52\%) show the most extensive changes. PD-L1 (7/32, 22\%) preserves structural disulfide cysteines. VHH72 (11/29, 38\%) changes H1 and H2 while largely preserving H3. anti-TNF (10/19, 53\%) undergoes extensive mutation but achieves minimal ipTM gain due to poor initial pose.}
\label{fig:logos}
\end{figure}

\subsection{Determinants of design success}

Analysis across the five targets reveals two principal determinants of design success:

\textbf{CDR length.} The two improved targets have $\geq 22$ CDR positions (Ty1: 22, PD-L1: 32). anti-TNF has only 19 positions, providing insufficient degrees of freedom. Longer CDRs---particularly the elongated H3 loops characteristic of nanobodies---provide both a larger optimization surface and greater conformational flexibility.

\textbf{Initial pose quality.} VHH72 and VHH3, with WT ipTM $>$ \num{0.65} and predicted poses within \SI{5}{\AA} of crystal, are already near-optimal. anti-TNF, with a \SI{44}{\AA} pose error, cannot be rescued. Ty1 and PD-L1 occupy the ``sweet spot'': weak WT ipTM but correct pose basin (CDRs positioned near the epitope), enabling substantial improvement through CDR optimization alone. This suggests a practical two-stage workflow: (1) screen framework sequences to identify those in the correct pose basin, then (2) optimize CDRs of promising frameworks.

\begin{figure}[H]
\centering
\includegraphics[width=0.49\textwidth]{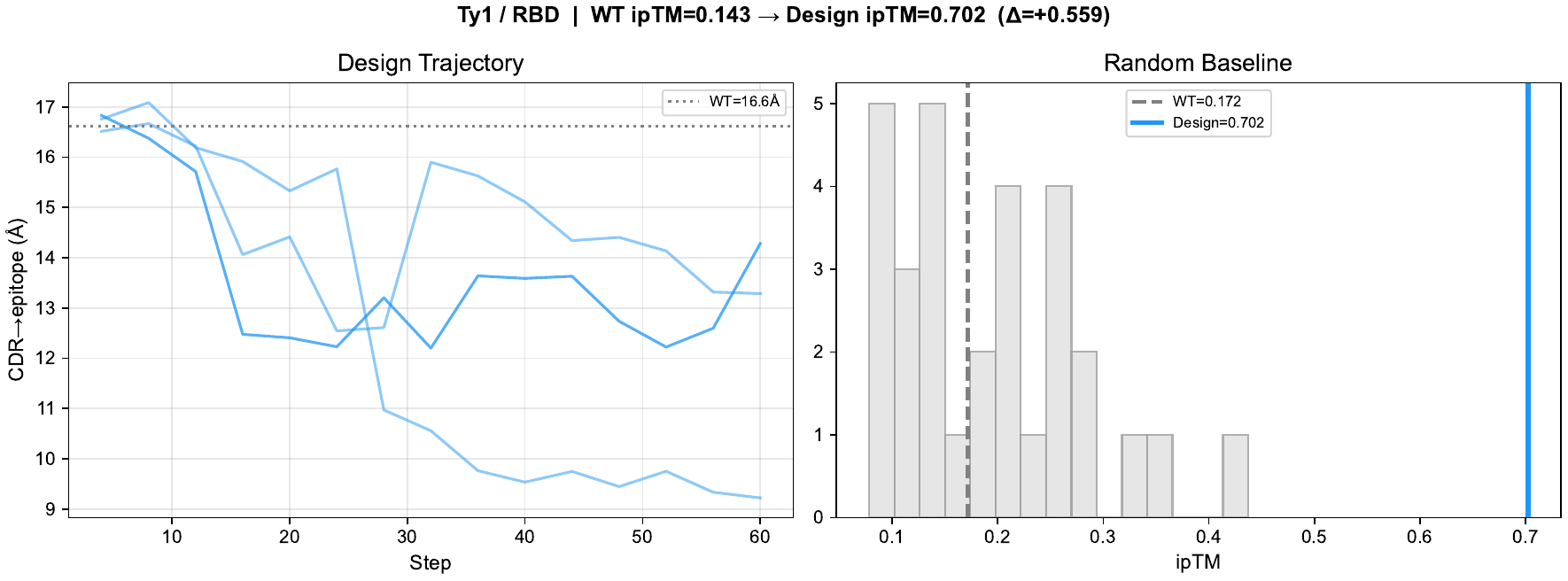}\hfill
\includegraphics[width=0.49\textwidth]{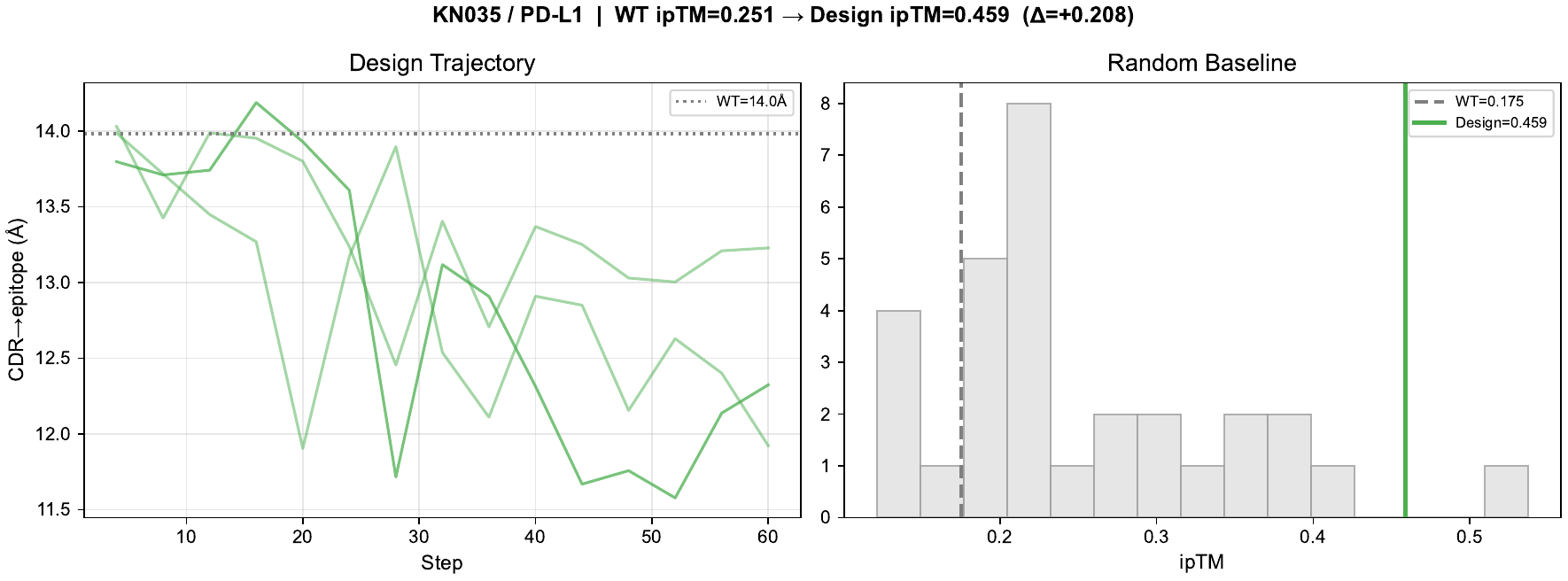}\\[4pt]
\includegraphics[width=0.49\textwidth]{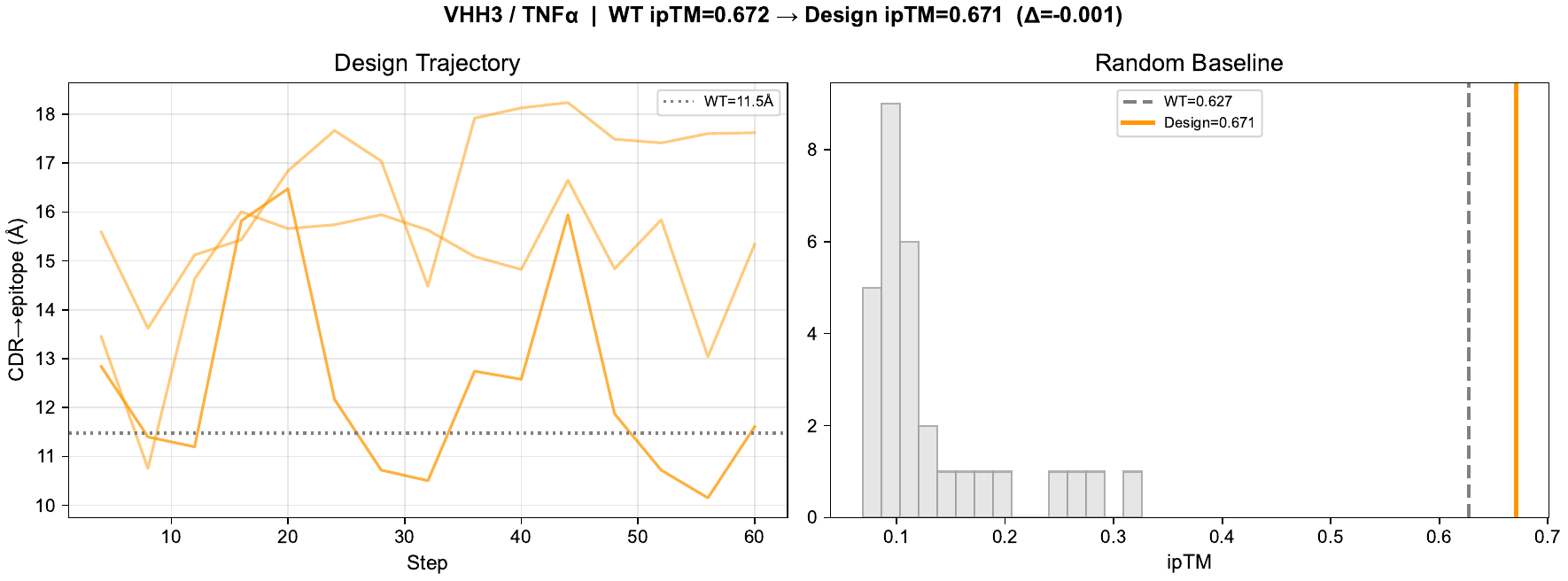}\hfill
\includegraphics[width=0.49\textwidth]{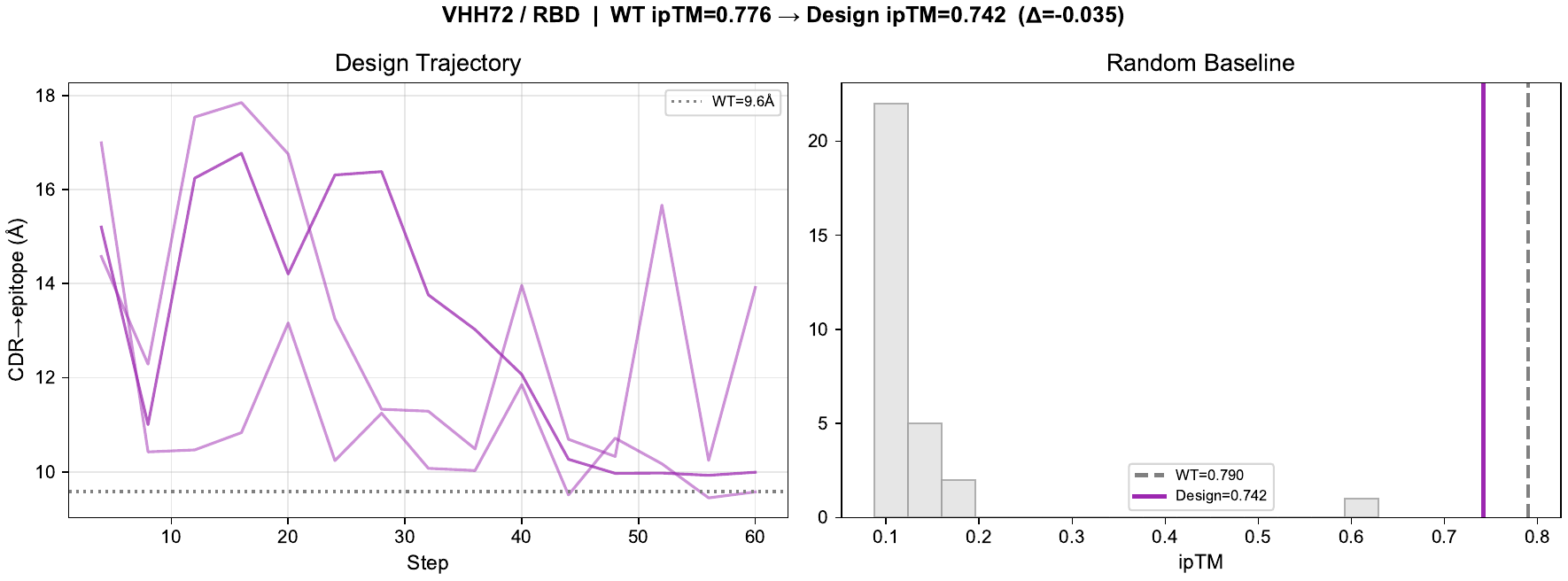}\\[4pt]
\centering
\includegraphics[width=0.49\textwidth]{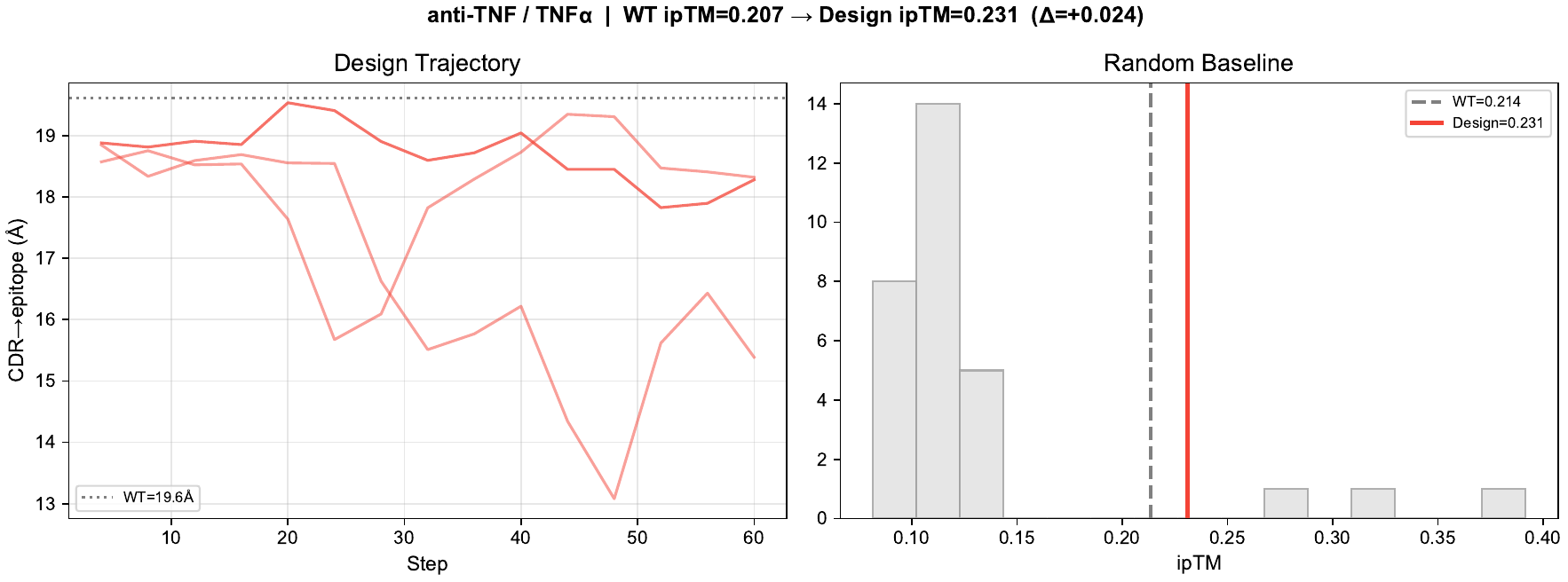}
\caption{\textbf{Extended Data Figure 1: Per-target design trajectories and random baseline distributions.}
For each target, left panel: CDR-to-epitope distance vs.\ optimization step (three independent seeds), right panel: random CDR baseline ipTM distribution with WT and design marked. (a) Ty1/RBD---best design achieves \SI{9.2}{\AA} cdr$\to$epi and ipTM 0.702, 5.7$\sigma$ above random. (b) KN035/PD-L1---design achieves ipTM 0.459, 2.2$\sigma$ above random, limited by disulfide constraint. (c) VHH3/TNF$\alpha$---strong WT binder, design preserved ipTM. (d) VHH72/RBD---strong WT binder, design preserved ipTM. (e) anti-TNF/TNF$\alpha$---marginal improvement despite extensive CDR mutation; poor initial pose limits success.}
\label{fig:per_target}
\end{figure}

\section{Discussion}

We have demonstrated that epitope-targeted nanobody CDR design can be performed rapidly---\textbf{approximately 10--20 minutes per target on a high-end personal workstation}---through gradient-based optimization of the ESMFold2 distogram. Our method achieves statistically significant ipTM improvements on weak binders (up to +\num{0.559} for Ty1/RBD), preserves strong binders, requires no experimental screening or explicit docking, and provides multiple diverse candidate CDR sequences.

\subsection{Relationship to the official ESMFold2 binder design code}

The official ESMFold2 codebase includes a binder design module (\texttt{binder\_design.py})\cite{lin2023} that introduced gradient-guided sequence optimization through the ESMFold2 distogram. We gratefully build upon this foundation. The official code performs general binder design---optimizing a sequence to contact a target at any interface---and does not support epitope-level targeting, structure priors for pose anchoring, or CDR-restricted optimization. EasyNano extends this framework with two capabilities essential for practical epitope-targeted nanobody design: (1) a user-specified epitope targeting mechanism via a dedicated CDR-to-epitope distance loss, which directs optimization toward specific target residues rather than any interface; and (2) a Full ESMFold2 CA-coordinate structure prior that anchors the framework pose during CDR optimization, preventing pose drift and enabling de novo design scenarios where no complex crystal structure exists. The core principle---backpropagating structural signals through the ESMFold2 distogram---is shared; EasyNano specializes this principle for the epitope-targeted nanobody CDR design problem and demonstrates its practical effectiveness across six diverse targets.

\subsection{Comparison with other computational methods}

EasyNano differs fundamentally from existing methods (Table~\ref{tab:comparison}): inverse folding\cite{dauparas2022,hsu2022} requires a pre-specified backbone; RFdiffusion\cite{watson2023} generates backbones then sequences via ProteinMPNN; DiffAb\cite{luo2022} and dyMEAN\cite{kong2023} use diffusion or equivariant networks requiring GPU clusters. In contrast, our method directly optimizes CDR sequences over a continuous relaxation, using a single ESMFold2-Fast forward pass per gradient step---\textbf{60 forward/backward passes of a 721M-parameter model, tractable on a single high-end personal workstation}. This speed enables rapid iteration and high-throughput framework variant screening.

\begin{table}[H]
\centering
\caption{\textbf{Comparison with existing computational binder design methods.}}
\label{tab:comparison}
\begin{tabular}{@{}lcccc@{}}
\toprule
Method & Epitope-specific & GPU time & Differentiable & No backbone required \\
\midrule
ProteinMPNN\cite{dauparas2022} & No & $<$1 min & No & No \\
ESM-IF\cite{hsu2022} & No & $<$1 min & No & No \\
RFdiffusion\cite{watson2023} & Partial & hours--days & No & Yes \\
DiffAb\cite{luo2022} & Yes & hours & Yes & Yes \\
dyMEAN\cite{kong2023} & Yes & hours & Yes & Yes \\
\textbf{EasyNano (this work)} & \textbf{Yes} & \textbf{$\sim$15 min} & \textbf{Yes} & \textbf{Yes} \\
\bottomrule
\end{tabular}
\end{table}

The wild-type logit bias ($\beta$) emerged as a critical practical parameter. Because epitope gradient signals are weak early in optimization (when CDRs are far from the epitope and the distogram is uncertain), the initialization strongly influences the optimization trajectory. Too large a WT bias ($\beta \ge 5.0$) prevents CDR mutation entirely; too little ($\beta \le 1.0$) permits chaotic drift before meaningful gradient signals develop. The optimal $\beta = 2.0$ provides sufficient WT preference to maintain physically reasonable sequences during early exploration while remaining weak enough for epitope-driven gradients to overcome. This balance may require per-target adjustment for unusual CDR compositions.

\subsection{ipTM as an evaluation metric and the proxy optimization paradigm}

The ipTM score has emerged as a leading computational metric for predicting antibody-antigen binding, showing strong correlation with experimental binding data in recent benchmarks\cite{lin2023,ruffolo2023}. However, direct ipTM optimization remains challenging because computing a single ipTM value requires a full-model forward pass with confidence head ($\sim$\SI{30}{\second}), making gradient-based optimization over hundreds of steps infeasible on a high-end personal workstation. EasyNano therefore optimizes a distogram-based proxy loss---a weighted combination of epitope proximity, structural contact, and pose preservation terms---and evaluates ipTM post hoc on top candidates. This indirect strategy works well in practice (ipTM improvements of up to +\num{0.559} for Ty1/RBD) but does not guarantee ipTM improvement in all cases: the Fast model's distogram signal can diverge from the full model's ipTM (Supplementary Table S2), particularly for frameworks with poor initial poses. We note that optimizing antibody ipTM directly remains an open challenge, and EasyNano's proxy-based approach represents a pragmatic step toward addressing it---one that achieves useful results while keeping the computational cost tractable for routine use.

Several limitations should be acknowledged. First, the method inherits ESMFold2's pose accuracy: if the framework's predicted pose is incorrect (as with anti-TNF, \SI{44}{\AA} from crystal), CDR optimization alone cannot compensate. Incorporating framework micro-tuning---single-residue mutations at framework positions flanking CDRs, which we have shown can shift the pose basin\cite{hu2026esmfold2}---may address this. Second, the Fast model's ipTM values are poorly calibrated (Fast-to-Full cdr$\to$epitope discrepancies of up to \SI{7}{\AA}), requiring full model evaluation for final ranking. Third, we have not yet experimentally validated designed sequences; binding must be confirmed by SPR or BLI. Fourth, the method currently optimizes only CDR sequences, not CDR lengths.

Future work should pursue: (1) experimental validation of top Ty1/RBD and PD-L1 designs; (2) incorporating ipTM as a direct loss term; (3) joint CDR length and sequence optimization; (4) framework residue co-optimization; (5) extension to multi-specificity design; and (6) application to other programmable scaffolds such as DARPins\cite{pluckthun2015} and monobodies\cite{koide2012}.

In conclusion, differentiable distogram optimization provides a fast, epitope-specific, and computationally efficient approach to nanobody CDR design. We anticipate applications in rapid prototyping of nanobody candidates against therapeutic targets, and expect the general principle---using structure prediction models as differentiable oracles for sequence optimization---to generalize to other protein design tasks.

\section{Methods}

\subsection{Target preparation}
Five target-framework pairs were selected from the PDB: Ty1 nanobody with SARS-CoV-2 spike (6ZXN, chains D/A), KN035 with PD-L1 (5JDS, B/A), VHH72 with SARS-CoV-2 RBD (6WAQ, A/B), anti-TNF VHH with TNF$\alpha$ (5M2J, D/A), and VHH3 with TNF$\alpha$ trimer (5M2M, D/B). For 6ZXN (spike trimer, \num{1060} residues per chain), the RBD domain (residues 330--535, \num{204} residues) was extracted. Epitope residues were target residues within \SI{8}{\AA} C$\alpha$--C$\alpha$ of any nanobody heavy atom, identified via Biotite\cite{kunzmann2023}. CDR positions used Chothia numbering via abnumber\cite{al1997}.

\subsection{Structure prior}
Full ESMFold2 (1.3B)\cite{lin2023} with 3 recycling loops, 14 sampling steps, 4 diffusion samples. CA coordinates (averaged across diffusion samples) were discretized into 64 distance bins (\SIrange{2}{22}{\AA}, tolerance \SI{2.5}{\AA}).

\subsection{Differentiable design loop}
CDR residue logits $\mathbf{L} \in \mathbb{R}^{n_\text{CDR} \times 20}$ initialized Gaussian ($\sigma=0.5$) + WT bias $\beta=2.0$. Temperature: $T_t = 0.1 + 0.9 \cdot \frac{1}{2}[1 + \cos(\pi t/60)]$. ESMFold2-Fast (721M, 1 loop, 5 sampling steps). Adam optimizer (lr=\num{0.05}). Framework positions pinned after each step. Sixty optimization steps.

\subsection{Evaluation}
Full ESMFold2 (3 loops, 14 sampling steps, 1 diffusion sample, confidence head enabled). Three independent seeds per target. Random CDR baselines: $n=30$ per target, amino acids sampled from natural frequencies (Cys excluded). Kabsch RMSD: predicted target CA aligned to crystal target CA; binder RMSD computed under same transform.

\subsection{Compute}
Apple Mac Studio M3 Ultra (Apple Silicon, 256 GB unified memory, MPS backend). ESMFold2 models in float32. MPS fallback enabled for unsupported operations. Design loop: $\sim$10--17 s/step depending on complex size (254--322 residues). Full pipeline per target (3 design seeds + 30 random baselines): approximately 2.0--3.5 hours.

\section*{Data availability}
All PDB structures from RCSB PDB (https://www.rcsb.org/): 6ZXN, 5JDS, 6WAQ, 5M2J, 5M2M. ESMFold2 and ESMC weights from EvolutionaryScale (https://github.com/evolutionaryscale/esm). Source data for all figures and tables are provided as Supplementary Data.

\section*{Code availability}
EasyNano is available at https://github.com/[organization]/EasyNano under MIT license. Frozen version archived at Zenodo (DOI to be provided).

\section*{Acknowledgements}
We thank EvolutionaryScale for releasing ESMFold2 model weights. Supported by Qilu University of Technology (Shandong Academy of Sciences) and Shandong Provincial Hospital.

\section*{Author contributions}
Y.H. conceived the study, developed the methodology, wrote all computational pipelines, performed experiments, and drafted the manuscript. W.C. contributed to data collection and computational experiments. J.W. contributed to target preparation, epitope analysis, and data interpretation. Y.L. provided clinical domain expertise, target selection guidance, and contributed to the manuscript. All authors reviewed and approved the final manuscript.

\section*{Competing interests}
The authors declare no competing interests.



\begin{thebibliography}{99}

\bibitem{muyldermans2013} Muyldermans, S. Nanobodies: natural single-domain antibodies. \textit{Annu. Rev. Biochem.} \textbf{82}, 775--797 (2013).
\bibitem{hamers1993} Hamers-Casterman, C. \textit{et al.} Naturally occurring antibodies devoid of light chains. \textit{Nature} \textbf{363}, 446--448 (1993).
\bibitem{jovcevska2020} Jov{\v{c}}evska, I. \& Muyldermans, S. The therapeutic potential of nanobodies. \textit{BioDrugs} \textbf{34}, 11--26 (2020).
\bibitem{scully2019} Scully, M. \textit{et al.} Caplacizumab treatment for acquired thrombotic thrombocytopenic purpura. \textit{N. Engl. J. Med.} \textbf{380}, 335--346 (2019).

\bibitem{norman2020} Norman, R.A. \textit{et al.} Computational approaches to therapeutic antibody design. \textit{Brief. Bioinform.} \textbf{21}, 1549--1567 (2020).

\bibitem{jumper2021} Jumper, J. \textit{et al.} Highly accurate protein structure prediction with AlphaFold. \textit{Nature} \textbf{596}, 583--589 (2021).
\bibitem{baek2021} Baek, M. \textit{et al.} Accurate prediction of protein structures and interactions using a three-track neural network. \textit{Science} \textbf{373}, 871--876 (2021).
\bibitem{lin2023} Lin, Z. \textit{et al.} Evolutionary-scale prediction of atomic-level protein structure with a language model. \textit{Science} \textbf{379}, 1123--1130 (2023).
\bibitem{abramson2024} Abramson, J. \textit{et al.} Accurate structure prediction of biomolecular interactions with AlphaFold 3. \textit{Nature} \textbf{630}, 493--500 (2024).

\bibitem{dauparas2022} Dauparas, J. \textit{et al.} Robust deep learning--based protein sequence design using ProteinMPNN. \textit{Science} \textbf{378}, 49--56 (2022).
\bibitem{hsu2022} Hsu, C. \textit{et al.} Learning inverse folding from millions of predicted structures. \textit{Proc. ICML} (2022).
\bibitem{anishchenko2021} Anishchenko, I. \textit{et al.} De novo protein design by deep network hallucination. \textit{Nature} \textbf{600}, 547--552 (2021).
\bibitem{watson2023} Watson, J.L. \textit{et al.} De novo design of protein structure and function with RFdiffusion. \textit{Nature} \textbf{620}, 1089--1100 (2023).
\bibitem{wang2022} Wang, J. \textit{et al.} Scaffolding protein functional sites using deep learning. \textit{Science} \textbf{377}, 387--394 (2022).

\bibitem{ruffolo2023} Ruffolo, J.A. \textit{et al.} Fast, accurate antibody structure prediction from deep learning on massive set of natural antibodies. \textit{Nat. Commun.} \textbf{14}, 2389 (2023).
\bibitem{ruffolo2021} Ruffolo, J.A., Gray, J.J. \& Sulam, J. Deciphering antibody affinity maturation with language models and weakly supervised learning. \textit{arXiv} (2021).
\bibitem{olsen2022} Olsen, T.H., Moal, I.H. \& Deane, C.M. AbLang: an antibody language model for completing antibody sequences. \textit{Bioinform. Adv.} \textbf{2}, vbac046 (2022).
\bibitem{luo2022} Luo, S. \textit{et al.} Antigen-specific antibody design and optimization with diffusion-based generative models for protein structures. \textit{Adv. Neural Inf. Process. Syst.} \textbf{35}, 9754--9767 (2022).
\bibitem{kong2023} Kong, X., Huang, W. \& Liu, Y. End-to-end full-atom antibody design. \textit{Proc. ICML} (2023).
\bibitem{jin2022} Jin, W. \textit{et al.} Iterative refinement graph neural network for antibody sequence-structure co-design. \textit{Proc. ICLR} (2022).

\bibitem{hanke2020} Hanke, L. \textit{et al.} An alpaca nanobody neutralizes SARS-CoV-2 by blocking receptor interaction. \textit{Nat. Commun.} \textbf{11}, 4420 (2020).
\bibitem{wang2021} Wang, Y. \textit{et al.} A potent neutralizing nanobody against SARS-CoV-2. \textit{Nat. Commun.} \textbf{12}, 4003 (2021).
\bibitem{zhang2019} Zhang, F. \textit{et al.} Structural basis of a novel PD-L1 nanobody for immune checkpoint blockade. \textit{Cell Discov.} \textbf{5}, 43 (2019).

\bibitem{he2022} He, B.L. \textit{et al.} A protein language model for all domains of life. \textit{bioRxiv} (2022).
\bibitem{rives2021} Rives, A. \textit{et al.} Biological structure and function emerge from scaling unsupervised learning to 250 million protein sequences. \textit{Proc. Natl. Acad. Sci.} \textbf{118}, e2016239118 (2021).

\bibitem{liu2024} Liu, Y. \textit{et al.} De novo design of programmable protein interactions. \textit{Nat. Rev. Bioeng.} (2024).
\bibitem{pluckthun2015} Pl{\"u}ckthun, A. Designed ankyrin repeat proteins (DARPins). \textit{Annu. Rev. Pharmacol. Toxicol.} \textbf{55}, 489--511 (2015).
\bibitem{koide2012} Koide, A. \textit{et al.} Teaching an old scaffold new tricks: monobodies. \textit{J. Mol. Biol.} \textbf{415}, 393--405 (2012).

\bibitem{kunzmann2023} Kunzmann, P. \textit{et al.} Biotite: a unifying open source computational biology framework in Python. \textit{BMC Bioinform.} \textbf{24}, 346 (2023).
\bibitem{al1997} Al-Lazikani, B., Lesk, A.M. \& Chothia, C. Standard conformations for the canonical structures of immunoglobulins. \textit{J. Mol. Biol.} \textbf{273}, 927--948 (1997).

\bibitem{hu2026esmfold2} Hu, Y. \textit{et al.} ESMFold2 predicted complex pose is determined by framework sequence, not diffusion initialization. (this work, internal analysis).

\end{thebibliography}
\end{document}